\documentclass[twocolumn,showpacs,preprintnumbers]{revtex4}
\usepackage{graphicx}
\usepackage{dcolumn}
\usepackage{bm}
\usepackage{epsfig}
\usepackage{amsmath}
\usepackage{subfigure}

\begin{document}

\title{Tunable double optomechanically induced transparency in an
optomechanical system}
\author{Peng-Cheng Ma$^{1,2,3}$}
\author{Jian-Qi Zhang$^{2}$}
\email{changjianqi@gmail.com}
\author{Yin Xiao$^1$}
\author{Mang Feng$^{2}$}
\email{mangfeng@wipm.ac.cn}
\author{Zhi-Ming Zhang$^{1}$}
\email{zmzhang@scnu.edu.cn}
\address{$^1$Laboratory of Nanophotonic Functional Materials and Devices (SIPSE), and
Laboratory of Quantum Engineering and Quantum Materials, South China
Normal University, Guangzhou 510006, China\\
$^2$State Key Laboratory of Magnetic Resonance and Atomic and
Molecular Physics, Wuhan Institute of Physics and Mathematics,
Chinese Academy of Sciences, Wuhan 430071, China
\\ $^3$School of Physics
and Electronic Electrical Engineering, Huaiyin Normal University,
Huaian 223300, China}

\begin{abstract}
We study the dynamics of a driven optomechanical cavity coupled to a
charged nanomechanical resonator via Coulomb interaction, in which
the tunable double optomechanically induced transparency (OMIT) can
be observed from the output field at the probe frequency by
controlling the strength of the Coulomb interaction. We calculate
the splitting of the two transparency windows, which varies near
linearly with the Coulomb coupling strength in a robust way against
the cavity decay. Our double-OMIT is much different from the previously
mentioned double-EIT or double-OMIT, and might be applied to measure the Coulomb coupling
strength.
\end{abstract}
\maketitle

\section{Introduction}

Recently, significant theoretical and experimental efforts have been paid on
studying the characteristic and application of nanomechanical
resonators (NRs) \cite{1,2,3,4}. NRs own some important properties, such as phonon induced
transparency \cite{5}, phonon blockade \cite{6}, and high harmonic generation \cite{7}, and can be employed
in many applications as, for example, single photon source \cite{8},
single phonon source \cite{9}, biological sensor \cite{10}, quantum
information processing \cite{11}, and quantum metrology \cite{12, 13}.

In combination with an optical cavity, an NR turns to be an optomechanical system \cite{14, 15, 16, 17}, in
which the NR interacts with the cavity mode via the radiation pressure force and
enables observation of the NR-induced quantum mechanical behaviors from the
output light of the cavity. Until now, there have been a lot of theoretical predictions in
such optomechanical systems, for example, photon
blockade \cite{18}, Kerr effect \cite{19}, optomechanically induced
transparency (OMIT) \cite{20}, quantum information transfer \cite{21},
normal-mode splitting \cite{22}, and some of them have been demonstrated experimentally,
 such as, OMIT \cite{23,24,25,26}, slow light \cite{24}, frequency
transfer \cite{27}, and normal-mode splitting \cite{28}.

The present work is focused on the OMIT effect in the optomechanical
cavity. The OMIT is a kind of induced transparency caused by the
radiation pressure in an optomechanical system \cite{20, 23}, which
stands at the center of current studies for optomechanics. We have
noticed recent OMIT-relevant work on four-wave mixing \cite{29},
superluminal and ultraslow light propagation \cite{30,31}, quantum
router \cite{32}, and precision measurement of electrical charge
\cite{33}. On the other hand, double electromagnetic induced
transparency (EIT) \cite{34,35,36} is a hot topic over recent
years, which extends conventional EIT to the one with double
transparency windows, and discovers some new physics and
applications. This arises a question: what would happen in an OMIT
with two transparency windows (i.e., double-OMIT)? To the best of
our knowledge, there have been a few theoretical schemes
\cite{37,38,39} for the double-OMIT with different models, using a
nonlinear crystal or a qubit in an optomechanical cavity \cite{37,38},
and using a ring cavity with two movable mirrors \cite{39}. However, in
all the schemes mentioned above, the frequency of
the transparency light for the double-EIT/OMIT cannot be changed due to
the fixed coupling for splitting the transparency windows.

In this work, we demonstrate a tunable double-OMIT observable
in an optomechanical system, in which the two NRs are charged and
the two transparency windows are split due to the Coulomb interaction.
Specifically, our optomechanical system consists of an
optomechanical cavity and a NR outside, as sketched in Fig. 1, where
the NR of the optomechanical cavity (i.e., NR$_{1}$) not only
couples to the cavity field by the radiation pressure, but also
interacts with the NR outside the cavity (i.e., NR$_{2}$) through
the tunable Coulomb interaction, which can be controlled by the bias
voltages on the NRs.

Compared with the conventional OMIT with a single transparency
window\cite{23,24,25,26}, our scheme owns some favorable features:
(i) The two output lights with different frequencies are controlled
by a single driving light; (ii) Our scheme is robust to the cavity
decay, and the transparency windows are with narrow profiles; (iii)
We find that the two windows of the double-OMIT are apart near
linearly with respect to the Coulomb coupling strength. The feature
reminds us of a practical application of the double-OMIT for
precisely detecting the Coulomb coupling strength. In this context,
we have to emphasize that our proposal is essentially different from
the previous ideas \cite{38,39}, where the double-OMIT is caused by
the frequency difference between the two NRs and the frequencies of
the transparency lights are fixed. In contrast, our studied
double-OMIT can be observed even for two identical NRs, and the
frequency of the transparency light can be selected by tuning the
Coulomb coupling under a constant driving light.


This paper is structured as follows. In Sec. \makeatletter\@Roman{2}
\makeatother we present the model and the analytical expressions of
the optomechanical system and obtain the steady-state mean values.
Sec. \makeatletter\@Roman{3} \makeatother includes numerical
calculations for the double-OMIT based on recent experimental
parameters. The feasibility of precision measurement of the Coulomb
coupling strength between the two NRs is discussed in Sec.
\makeatletter\@Roman{4} and we also justify the robustness of our
approach against the cavity decay.   The last section is a brief
conclusion.

\section{the Model and the solutions}

\begin{figure}[tbp]
\centering
\includegraphics[width=8cm,height=4cm]{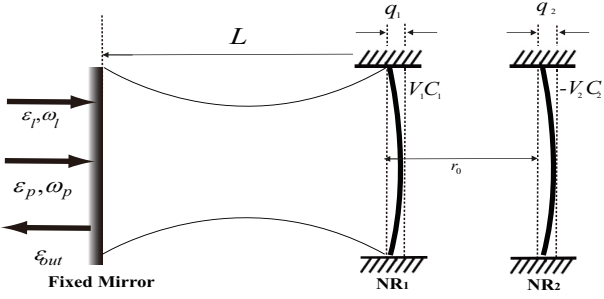}
\caption{Schematic diagram of the system. A high-quality
Fabry-P\'{e}rot cavity consists of a fixed mirror and a movable
mirror NR$_{1}$. NR$_{1}$ is charged by the bias gate voltage
$V_{1}$ and subject to the Coulomb force due to another charged
NR$_{2}$ with the bias gate voltage $-V_{2}$. The optomechanical
cavity of the length $L$ is driven by two light fields, one of which
is the pump field $\protect\varepsilon_{l}$ with frequency
$\protect\omega_{l}$ and the other of which is the probe field
$\protect\varepsilon_{p}$ with frequency $\protect\omega_{p}$.
The output field is represented by $\protect\varepsilon_{out}$.
$q_{1}$ and $q_{2}$ represent the small displacements of NR$_{1}$
and NR$_{2}$ from their equilibrium positions, with $r_{0}$ the
equilibrium distance between the two NRs.}
\end{figure}

For the system in Fig. 1, the Hamiltonian is given by,
\begin{eqnarray}
&&H_{whole}=\hbar \omega _{c}c^{\dag
}c+(\frac{p_{1}^{2}}{2m_{1}}+\frac{1}{2}
m_{1}\omega _{1}^{2}q_{1}^{2})  \notag \\
&&+(\frac{p_{2}^{2}}{2m_{2}}+\frac{1}{2}m_{2}\omega _{2}^{2}q_{2}^{2})-\hbar
gc^{\dag }cq_{1}+H_{C}  \notag \\
&&+i\hbar \varepsilon _{l}(c^{\dag }e^{-i\omega _{l}t}-h.c.)+i\hbar
(c^{\dag }\varepsilon _{p}e^{-i\omega _{p}t}-h.c.),
\end{eqnarray}
where the first term is for the single-mode cavity field with
frequency $\omega_{c}$ and annihilation (creation) operator $c\
(c^{\dag })$. The second (third) term describes the vibration of the
charged NR$_{1}$ (NR$_{2}$ ) with frequency $\omega_{1}$
($\omega_{2}$), effective mass $m_{1}$ ($ m_{2}$), position $q_{1}$
($q_{2}$) and momentum operator $p_{1}$ ($p_{2}$) \cite{32}.
NR$_{1}$ couples to the cavity field due to the radiation pressure
with the coupling strength $g=\frac{\omega_{c}}{L}$ with $L$ being
the cavity length.

The fifth term $H_{C}$ in Eq.(1) presents the Coulomb coupling
between the charged NR$_{1}$ and NR$_{2}$ \cite{40}, where the
NR$_{1}$ and NR$_{2}$ take the charges $C_{1}V_{1}$ and
$-C_{2}V_{2}$, with $C_{1}(C_{2})$ and $V_{1}(-V_{2})$ being the
capacitance and the voltage of the bias gate, respectively. So the
Coulomb coupling between NR$_{1} $ and NR$_{2}$ is given by
\begin{equation*}
H_{C}=\frac{-C_{1}V_{1}C_{2}V_{2}}{4\pi \varepsilon
_{0}|r_{0}+q_{1}-q_{2}|},
\end{equation*}%
where $r_{0}$ is the equilibrium distance between NR$_{1}$ and
NR$_{2}$, $q_{1}$ and $q_{2}$ represent the small displacements of
NR$_{1}$ and NR$_{2}$ from their equilibrium positions,
respectively. In the case of $r_{0}\gg q_{1},q_{2}$, with the second
order expansion, the Hamiltonian above is rewritten as
\begin{equation*}
H_{C}=\frac{-C_{1}V_{1}C_{2}V_{2}}{4\pi \varepsilon
_{0}r_{0}}[1-\frac{
q_{1}-q_{2}}{r_{0}}+(\frac{q_{1}-q_{2}}{r_{0}})^{2}],
\end{equation*}
where the linear term may be absorbed into the definition of the
equilibrium positions, and the quadratic term includes a
renormalization of the oscillation frequency for both NR$_{1}$ and
NR$_{2}$. This implies a reduced form
\begin{equation*}
H_{C}=\hbar \lambda q_{1}q_{2},
\end{equation*}
where $\lambda =\frac{C_{1}V_{1}C_{2}V_{2}}{2\pi \hbar \varepsilon
_{0}r_{0}^{3}}$ \cite{40, 41, 42}.

The last two terms in Eq. (1) describe the interactions between the
cavity field and the two input fields, respectively. The strong
(week) pump (probe) field owns the frequency $\omega_{l}$
($\omega_{p}$) and the amplitude $
\varepsilon_{l}=\sqrt{2\kappa\wp_{l}/\hbar\omega_{l}}$ ($\varepsilon
_{p}= \sqrt{2\kappa \wp _{p}/\hbar\omega _{p}}$), where $\wp_{l}$
($\wp _{p}$) is the power of the pump (probe) field and $\kappa$ is
the cavity decay rate.

In a frame rotating with the frequency $\omega_l$ of the pump field, the
Hamiltonian of the total system Eq.(1) can be rewritten as,
\begin{eqnarray}
&&H=\hbar\Delta_cc^\dag c +(\frac{p_1^2}{2m_1}+\frac{1}{2}m_1\omega_1^2q_1^2)
\notag \\
&&+(\frac{p_2^2}{2m_2}+\frac{1}{2}m_2\omega_2^2q_2^2) -\hbar gc^\dag cq_1
+\hbar \lambda q_1q_2  \notag \\
&&+i\hbar\varepsilon_l(c^\dag -c)
+i\hbar(c^\dag\varepsilon_pe^{-i\delta t}-h.c.),
\end{eqnarray}
where $\Delta_c=\omega_c-\omega_l$ is the detuning of the pump field
from the bare cavity, and $\delta=\omega_p-\omega_l$ is the detuning
of the probe field from the pump field.

Considering photon losses from the cavity and the Brownian noise
from the environment, we may describe the dynamics of the system
governed by Eq. (2) using following nonlinear quantum Langevin
equations \cite{32},
\begin{eqnarray}
&&\dot{q_{1}}=\frac{p_{1}}{m_{1}},  \notag  \label{6} \\
&&\dot{p_{1}}=-m_{1}\omega _{1}^{2}q_{1}-\hbar \lambda q_{2}+\hbar
gc^{\dag
}c-\gamma _{1}p_{1}+\sqrt{2\gamma _{1}}\xi _{1}(t),  \notag \\
&&\dot{q_{2}}=\frac{p_{2}}{m_{2}},  \notag \\
&&\dot{p_{2}}=-m_{2}\omega _{2}^{2}q_{2}-\hbar \lambda q_{1}-\gamma
_{2}p_{2}+\sqrt{2\gamma _{2}}\xi _{2}(t),  \notag \\
&&\dot{c}=-[\kappa +i(\Delta _{c}-gq_{1})]c+\varepsilon _{l}+\varepsilon
_{p}e^{-i\delta t}+\sqrt{2\kappa }c_{in},
\end{eqnarray}
where $\gamma _{1}$ and $\gamma _{2}$  are the decay rates for NR$
_{1}$ and NR$_{2}$, respectively. The quantum Brownian noise $\xi
_{1}$ $(\xi _{2})$ comes from the coupling between NR$_{1}$
(NR$_{2}$) and its own environment with zero mean value \cite{43}.
$c_{in}$ is the input vacuum noise operator with zero mean value
\cite{43}. Under the mean field approximation $\langle Qc\rangle
=\langle Q\rangle \langle c\rangle $ \cite{20}, the mean value
equations are given by
\begin{eqnarray}
&&\langle \dot{q_{1}}\rangle =\frac{\langle p_{1}\rangle }{m_{1}},  \notag
\label{7} \\
&&\langle \dot{p_{1}}\rangle =-m_{1}\omega _{1}^{2}\langle
q_{1}\rangle -\hbar \lambda \langle q_{2}\rangle +\hbar g\langle
c^{\dag }\rangle \langle
c\rangle -\gamma _{1}\langle p_{1}\rangle ,  \notag \\
&&\langle \dot{q_{2}}\rangle =\frac{\langle p_{2}\rangle }{m_{2}},  \notag \\
&&\langle \dot{p_{2}}\rangle =-m_{2}\omega _{2}^{2}\langle
q_{2}\rangle -\hbar \lambda \langle q_{1}\rangle -\gamma _{2}\langle
p_{2}\rangle ,
\notag \\
&&\langle \dot{c}\rangle =-[\kappa +i(\Delta _{c}-g\langle q_{1}\rangle
)]\langle c\rangle +\varepsilon _{l}+\varepsilon _{p}e^{-i\delta t},
\end{eqnarray}
which is a set of nonlinear equations and the steady-state response
in the frequency domain is composed of many frequency components. We
suppose the solution with the following form \cite{33}
\begin{eqnarray}
&&\langle q_{1}\rangle =q_{1s}+q_{1+}\varepsilon _{p}e^{-i\delta
t}+q_{1-}\varepsilon _{p}^{\ast }e^{i\delta t},  \notag  \label{8} \\
&&\langle p_{1}\rangle =p_{1s}+p_{1+}\varepsilon _{p}e^{-i\delta
t}+p_{1-}\varepsilon _{p}^{\ast }e^{i\delta t},  \notag \\
&&\langle q_{2}\rangle =q_{2s}+q_{2+}\varepsilon _{p}e^{-i\delta
t}+q_{2-}\varepsilon _{p}^{\ast }e^{i\delta t},  \notag \\
&&\langle p_{2}\rangle =p_{2s}+p_{2+}\varepsilon _{p}e^{-i\delta
t}+p_{2-}\varepsilon _{p}^{\ast }e^{i\delta t},  \notag \\
&&\langle c\rangle =c_{s}+c_{+}\varepsilon _{p}e^{-i\delta
t}+c_{-}\varepsilon _{p}^{\ast }e^{i\delta t},
\end{eqnarray}
where each quantity contains three items $O_{s}$, $O_{+}$, $O_{-}$
(with $ O\in \{q_{1},\ p_{1},\ q_{2},\ p_{2},\ c \}$), corresponding
to the responses at the frequencies $\omega _{l}$, $\omega _{p}$,
and $2\omega _{l}-\omega _{p}$, respectively \cite{44}. In the case
of $O_{s}\gg O_{\pm }$, Eq. (\ref{7}) can be solved by treating
$O_{\pm }$ as perturbations. After substituting Eq. (\ref{8}) into
Eq. (\ref{7}), and ignoring the second-order terms, we obtain the
steady-state mean values of the system as
\begin{eqnarray}
&&p_{1s}=p_{2s}=0,  \notag \\
&&q_{1s}=\frac{\hbar g|c_{s}|^{2}}{m_{1}\omega _{1}^{2}-\frac{\hbar
^{2}\lambda ^{2}}{m_{2}\omega _{2}^{2}}},  \notag \\
&&q_{2s}=\frac{\hbar \lambda q_{1s}}{-m_{2}\omega _{2}^{2}},  \notag \\
&&c_{s}=\frac{\varepsilon _{l}}{i\Delta +\kappa },  \notag \\
&&|c_{s}|^{2}=\frac{|\varepsilon_{l}|^{2}}{\Delta^{2}+\kappa^{2}},
\end{eqnarray}
with $\Delta =\Delta _{c}-gq_{1s}$, and
\begin{equation}
c_{+}=\frac{[\kappa -i(\Delta +\delta )][(\delta ^{2}-\omega
_{1}^{2}+i\delta \gamma _{1})-G]-2i\omega _{1}\beta }{[\Delta
^{2}-(\delta +i\kappa )^{2}][(\delta
^{2}-\omega_{1}^{2}+i\delta\gamma_{1})-G]+4\Delta \omega_{1}\beta },
\label{EIT}
\end{equation}
where $\beta =\frac{|c_{s}|^{2}\hbar g^{2}}{2m\omega_{1}}$ and
$G=\frac{ \hbar^2\lambda ^{2}}{m_1m_2(\delta
^{2}-\omega_{2}^{2}+i\delta\gamma_{2})}$. When there is no Coulomb
coupling $\lambda$ (i.e., $G=0$) between the two NRs, Eq. (\ref
{EIT}) is reduced to Eq. (5) in Ref. \cite{20}. However, different
from the output field in Ref. \cite{20} involving a single center
frequency for the single-mode OMIT, there are two centers with
different frequencies in our scheme due to the Coulomb interaction.
As a result, under the actions of the radiation pressure and the
probe light, two OMITs with different centers are reconstructed,
yielding the double-OMIT.

Making use of the input-output relation of the cavity \cite{45}
\begin{equation*}
\varepsilon _{out}(t)+\varepsilon _{p}e^{-i\delta t}+\varepsilon
_{l}=2\kappa \langle c\rangle ,
\end{equation*}
and
\begin{equation*}
\varepsilon _{out}(t)=\varepsilon _{outs}+\varepsilon _{out+}\varepsilon
_{p}e^{-i\delta t}+\varepsilon _{out-}\varepsilon _{p}^{\ast }e^{i\delta t},
\end{equation*}
we obtain
\begin{equation*}
\varepsilon _{out+}=2\kappa c_{+}-1,
\end{equation*}
which can be measured by homodyne technique \cite{45}. This output
light $\varepsilon _{out+}$ is of the same frequency $\omega _{p}$
as the probe field. Defining
\begin{equation}
\varepsilon _{T}=\varepsilon _{out+}+1=2\kappa c_{+},  \label{eq001}
\end{equation}
yields the real and imaginary parts, with $Re[\varepsilon _{T}]$ and $%
Im[\varepsilon _{T}]$, representing the absorption and dispersion of the
optomechanical system, respectively \cite{20}.

\section{double-OMIT in the output field}

\begin{figure}[tbp]
\centering
\includegraphics[width=8cm,height=4.9cm]{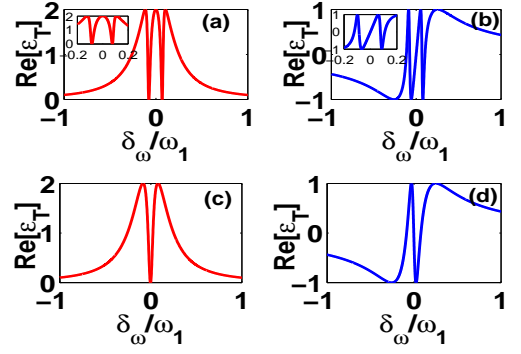}
\caption{(Color online) (a) The absorption
$Re[\protect\varepsilon_T]$ and (b) the dispersion
$Im[\protect\varepsilon_T]$ as functions of $\delta_\omega/\protect
\omega_1$ under the Coulomb interaction. (c) The absorption
$Re[\protect\varepsilon_T]$ and (d) the dispersion
$Im[\protect\varepsilon_T]$ as functions of
$\delta_\omega/\protect\omega_1$ in the absence of the Coulomb
interaction. $\delta_\omega= \protect\delta-\protect\omega_1$ is the
detuning from the central line of the sideband,
$\protect\lambda_l=$1064 nm, $L=$25 mm, $\protect\omega_1=\protect
\omega_2=2\protect\pi\times947\times10^3$ Hz, the quality factor
$Q_1=\frac{\omega_1}{\gamma_1}(=Q_2=\frac{\omega_2}{\gamma_2})=6700$,
$m_1=m_2=145$ ng,
$\protect\kappa=2\protect\pi\times215\times10^3$Hz, $\wp_l=2$ mW,
and $\protect\lambda=8\times10^{35}$ Hz/m$^2$ \cite{27}. }
\end{figure}

We present below the feasibility of the tunable double-OMIT in the
optomechanical  system, and the relationship between the double-OMIT
and the Coulomb interaction between the two NRs. As an estimate for
Eq.(\ref{eq001}), we employ the parameters from the recent
experiment \cite{27} in the observation of the normal-mode
splitting. For simplicity, we first consider two identical NRs in
our numerics, which is not essentially different in physics from the
case of two different NRs. We will also treat the different NRs
later.

As shown in Fig. 2, the absorption $Re[\varepsilon_{T}]$ and
dispersion $Im[\varepsilon _{T}]$ of the output field are plotted as
functions of $\delta_\omega/\omega_{1}=(\delta -\omega
_{1})/\omega_{1}$ for different Coulomb couplings. We may find that
the output lights for the probe field behave from the double-OMIT to
the single-mode OMIT with diminishing Coulomb coupling. The physics
behind the double-OMIT phenomenon can be understood from the
interference \cite{23,46} and the level configuration in Fig. 3.

The OMIT originates from the radiation pressure coupling an optical
mode to a mechanical mode. The simultaneous presence of the pump and
probe fields generates a radiation-pressure force oscillating at the
frequency difference $\delta=\omega_p-\omega_l$. If this frequency
difference is close to the resonance frequency $\omega_1$ of NR$_1$,
the mechanical mode starts to oscillate coherently. This in turn
gives rise to the Stokes- and anti-Stokes scattering of light from
the strong pump field. If the system is operated within the
resolved-sideband regime $\kappa\ll\omega_1$, the Stokes scattering
is strongly suppressed since it is highly off-resonant with the
optical cavity. We can therefore assume that only an anti-Stokes
field with frequency $\omega_p=\omega_l+\omega_1$ builds up inside
the cavity.  However, since this field is degenerate with the probe
field sent into the cavity, the two fields interfere destructively,
suppressing the case of a single transparency window for the output
beam. Thus the OMIT occurs.
As it depends on quantum interference, the OMIT is sensitive to
phase disturbances.   The coupling between NR$_{1}$ and NR$_{2}$
not only adds a fourth level, as shown in Fig. 3, but also breaks
down the symmetry of the OMIT interference, and thereby produces a
spectrally sharp bright resonance within the OMIT line shape. Then
the single OMIT transparency window is split into two transparency
windows, which yields the double-OMIT.

\begin{figure}[tbp]
\centering
\includegraphics[width=6cm,height=4cm]{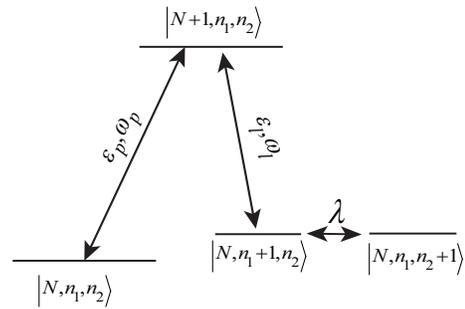}
\caption{ Schematic of the energy-level diagram in the cavity
optomechanical system, where $|N\rangle$, $|n_{1}\rangle$ and
$|n_{2}\rangle$ denote the number states of the cavity photon, and
NR$_{1}$ and NR$ _{2}$ phonons, respectively.
$|N,n_{1},n_{2}\rangle\longleftrightarrow |N+1,n_{1},n_{2}\rangle$
transition changes the cavity field, $
|N+1,n_{1},n_{2}\rangle\longleftrightarrow |N,n_{1}+1,n_{2}\rangle$
transition is caused by the radiation pressure coupling, and $
|N,n_{1}+1,n_{2}\rangle\longleftrightarrow |N,n_{1},n_{2}+1\rangle$
transition is induced by the Coulomb coupling \protect\cite{23,46}.}
\end{figure}

\section{Measurement of the coupling strength between NR$_1$ and NR$_2$}

\begin{figure}[tbp]
\centering
\includegraphics[width=8cm,height=5cm]{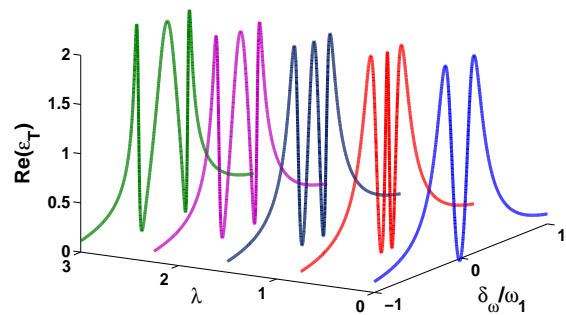}
\caption{(Color online) The absorption $Re[\protect\varepsilon_T]$
as functions of $\delta_\omega/\protect\omega_1$ and $\protect
\lambda$ (units of $\protect\lambda_0=8\times10^{35}$
Hz/m$^2$). Other parameters take the same values as in Fig. 2.}
\end{figure}

To further explore the characteristic of the tunable double-OMIT, we
plot the absorption $Re[\varepsilon_{T}]$ as functions of
$\delta_\omega/\omega_{1}$ and $\lambda$. One can find from Fig. 4
that only a single transparency window appears at $\delta_\omega=0\
(\delta =\omega _{1})$ in the absence of the Coulomb coupling, and
the single transparency window is split into two transparency
windows once the Coulomb coupling is present. The two transparency
windows are more and more apart with the increase of $\lambda$. The
two minima of the absorption in Fig. 4 can be evaluated by
\begin{equation}
\frac{dRe[\varepsilon
_{T}]}{d\delta_\omega}|_{\delta_\omega=\delta_{\omega+}}=0, \ \ \ \
\ \ \ \  \frac{dRe[\varepsilon
_{T}]}{d\delta_\omega}|_{\delta_\omega=\delta_{\omega-}}=0,
\end{equation}
where the detunings $\delta_{\omega+}$ and $\delta_{\omega-}$ are
the points with absorption minima. So the separation of the minima
is $d=|\delta_{\omega+}-\delta_{\omega-}|$, as plotted in Fig. 5,
where the almost linear increase of $d$ with $\lambda$ within the
regime $\lambda=\{0, 15\lambda_{0}\}$ reminds us of the possibility
to detect the Coulomb coupling strength between NR$_{1}$ and
NR$_{2}$ by measuring the separation $d$ in the absorption spectrum
$Re[\varepsilon_{T}]$ of the output field. From Fig. 5, one can
calculate the measuring sensitivity by $\frac{\partial
d}{\partial\lambda}$ on the order of $10^{-31}$m$^2$. Considering a
Coulomb coupling change $\Delta F$ due to a slight deviation
$q_{1}$, we have $\Delta F=\hbar\lambda q_{1}$. Provided
$q_{1}\approx$0.1 nm, we may assess $\partial\Delta F/\partial d$ to
be of the order of $10^{-13}$N/Hz, implying the possible precision
of measuring $\Delta F$ decided by the resolution of $d$ in the
absorption spectrum.

\begin{figure}[tbp]
\centering
\includegraphics[width=8cm,height=5cm]{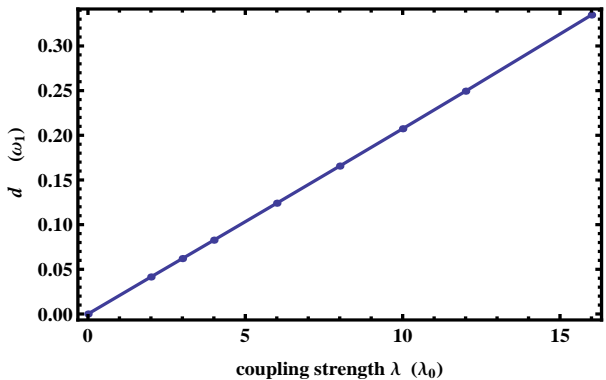}
\caption{ The separation $d$  (units of $\omega_1$) between the
two minima in the absorption spectrum as a function of the coupling
strength $\protect\lambda$ (units of $\protect\lambda_0=8\times10^{35}$Hz/m$^2$). Other parameters take
the same values as in Fig. 2.}
\end{figure}

\begin{figure}[tbp]
\centering
\includegraphics[width=8cm,height=5cm]{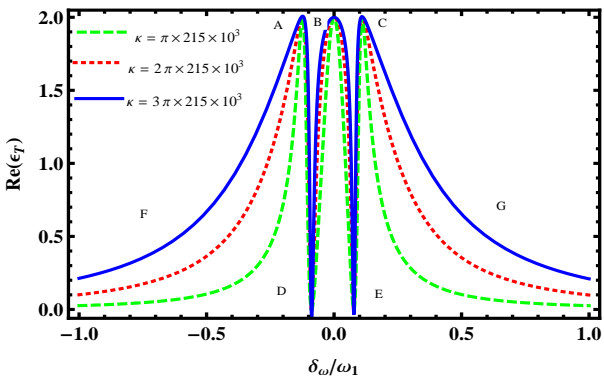}
\caption{(Color online) The absorption $Re[\protect\varepsilon_T]$
as a function of $\delta_\omega/\protect\omega_1$ with different
cavity decay rates,
$\protect\kappa=\protect\pi\times215\times10^3$Hz (green dashed
line), $\protect\kappa=2\protect\pi\times215\times10^3 $Hz (red
dotted line), $\protect\kappa=3\protect\pi\times215\times10^3 $Hz
(blue solid line). Other parameters take the same values as in Fig.
2.}
\end{figure}

Fig. 6 presents the variation of the absorption
$Re(\varepsilon_{T})$ with respect to $\delta_\omega/\omega_{1}$ for
different cavity decay rates, where the maxima (i.e., the points A,
B and C) and the minima (i.e., the points D and E) of the curves
remain unchanged in the parameter changes,  but the profiles of the
transparency window become narrower and sharper with the cavity
decay rate  $\kappa$ increasing. Provided a fixed driving light, the
bigger cavity decay will disturb the radiation pressure and makes it
less precise in detecting the strength of the radiation pressure,
which is reflected in Fig. 6 that the parts of the spectrum, FAD and
ECG, become wider and wider with $\kappa$ increasing. In contrast,
the other parts of the spectrum, ADB and BEC, turn to be narrower
and narrower, implying more precision in detecting the Coulomb
interaction. In comparison with the previous proposals \cite{12,13}
for detecting coupling strength, our double-OMIT can provide a more
effective and suitable method to achieve a precision measurement due
to the robustness against $\kappa$ and the narrower profiles in the
output light fields.

The robustness of our scheme can be understood as follows.  When the
Coulomb interaction and the driving light are fixed, the equilibrium
position is decided by the strain of the NR. With the increase of
the cavity decay rate, the radiation pressure in the optomechanical
system decreases, while the NR will acquire a larger displacement to
provide a larger strain for compensating the reduced radiation
pressure. The spectrum of the output becomes narrower for the larger
displacement of the NR. In this way, our scheme can be robust
against the cavity decay rate.

Moreover, for two different NRs, the results will be slightly
different from those above for identical NRs. Considering
$\omega_1\neq\omega_2$ in the calculation, we have plotted in Fig. 7
the absorption of the double-OMIT with larger separations of the
minima in comparison with the identical NR case. It implies a more
sensitivity to the coupling strength $\lambda$ in the case of two
different NRs. With respect to the $\omega_1=\omega_2$ case, the absorption
curves move rightward (leftward) in the case of $\omega_1>\omega_2$ ($\omega_1<\omega_2$). 
The enhancement of the sensitivity to the Coulomb force 
can be calculated by Eq. (9) and $d=|\delta_{\omega+}-\delta_{\omega-}|$, 
and is exemplified in Fig. 7 as 1.139 (1.529) times using $\omega_2=1.1\omega_1$
($\omega_2=0.8\omega_1$).

\begin{figure}[tbp]
\centering
\includegraphics[width=8cm,height=5cm]{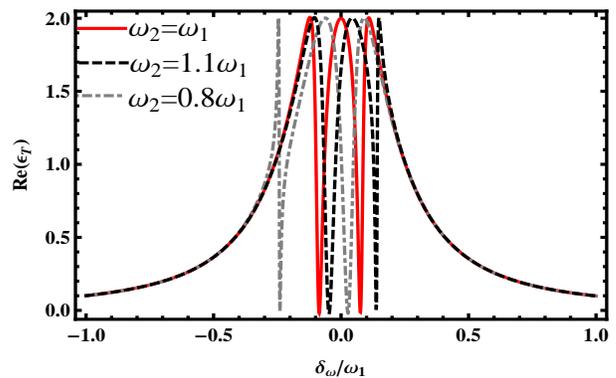}
\caption{(Color online) The absorption $Re[\protect\varepsilon_T]$
as a function of $\delta_\omega/\protect\omega_1$ for identical and
different NR frequencies. Other parameters take the same values as
in Fig. 2.}
\end{figure}

We have to mention that the robustness discussed above is limited
within the resolved regime ($\kappa <\omega_{1}$) where the
double-OMIT works. In contrast, the unresolved regime ($\kappa
>\omega_{1}$) blurs the sideband transitions, which makes the
quantum interference unavailable.

\section{Conclusion}


In conclusion, we have demonstrated the feasibility of the tunable
double-OMIT in the optomechanical system under the Coulomb
interaction between two charged NRs. To our knowledge, this is the
first proposal for the tunable double-OMIT in the optomechanical
system. Although our proposal is in principle extendable to other
interactions, such as the dipole-dipole coupling, the Coulomb
coupling, as a long-range interaction, is easier to control, and
thereby more practical. We have to emphasize that our double-OMIT is
neither a simple extension of the conventional OMIT nor a simple
transformation of the previously discussed double-EIT. Due to narrow
profiles of the transparency windows and robustness against
dissipation, the double OMIT might be employed for precisely
detecting the Coulomb coupling strength. Therefore, we argue that
our scheme have paved a new avenue towards the study of the OMIT
with more transparency windows as well as the relevant application.

\section*{ACKNOWLEDGMENTS}
PCM thanks Lei-Lei Yan and Wan-Lu Song for their helps in
the numerical simulation. JQZ thanks Yong Li for the helpful discussion.
This work was supported by the "973" Program (Grant No. 2011CBA00200,
No. 2012CB922102 and No. 2013CB921804), the Major Research Plan of the NSFC
(Grant No. 91121023), the NSFC (Grants No. 61378012, No. 60978009,
No. 11274352 and No. 11304366), the SRFDPHEC (Grant
No.20124407110009),  the PCSIRT (Grant No. IRT1243), China
Postdoctoral Science Foundation (Grant No. 2013M531771 and No.
2014T70760), Natural Science Funding for Colleges and Universities in
Jiangsu Province (Grant No. 12KJD140002), and Program for Excellent
Talents of Huaiyin Normal University(No. 11HSQNZ07).

\end{document}